\documentclass{PoS}
\usepackage{epsf}

\newcommand{\half}{\mbox{\small $\frac{1}{2}$}}          
\newcommand{\third}{\mbox{\small $\frac{1}{3}$}}         
\newcommand{\msbar}{\mbox{\tiny $\overline{MS}$}}        
\newcommand{\rimom}{\mbox{\tiny $R\!I\!\!-\!\!M\!O\!M$}}
\newcommand{\rgi}{\mbox{\tiny $R\!G\!I$}}                
\newcommand{\M}{\mbox{\tiny $({\cal M})$}}               

\newcommand{\Dr}{\stackrel{\rightarrow}{\cal D}}
\newcommand{\Dl}{\stackrel{\leftarrow}{\cal D}}
\newcommand{\Dlr}{\stackrel{\leftrightarrow}{\cal D}}

\def\lsim{\mathrel{\rlap{\lower4pt\hbox{\hskip1pt$\sim$}}
    \raise1pt\hbox{$<$}}}                
\def\gsim{\mathrel{\rlap{\lower4pt\hbox{\hskip1pt$\sim$}}
    \raise1pt\hbox{$>$}}}                
\title{
\vspace*{-2cm}
\begin{minipage}{\textwidth}
\begin{flushright}
\texttt{\footnotesize
PoS(LAT2007)144    \\%
DESY 07-177        \\%
Edinburgh 2007/28  \\%
Liverpool LTH 776  \\%
}
\end{flushright}
\end{minipage}\\[15pt]
\vspace*{+2cm}
       Distribution Amplitudes of Vector Mesons}

\ShortTitle{Distribution Amplitudes of Vector Mesons}

\author{V.~M.~Braun$^{a}$, D.~Br\"ommel$^{b}$, M.~G\"ockeler$^{a}$,
        \speaker{R.~Horsley}$^{\,c}$, Y.~Nakamura$^d$,
        H.~Perlt$^{e}$, D.~Pleiter$^d$, P.~E.~L.~Rakow$^f$,
        A.~Sch\"afer$^{a}$, G.~Schierholz$^{bd}$, A.~Schiller$^{e}$,
        W.~Schroers$^d$, T.~Streuer$^g$, H.~St\"uben$^h$
        and J.~M.~Zanotti$^c$ \\
        \llap{$^a$} Institut f\"ur Theoretische Physik,
                    Universit\"at Regensburg,
                    93040 Regensburg, Germany \\
        \llap{$^b$} Deutsches Elektronen-Synchrotron DESY,
                    22603 Hamburg, Germany \\
        \llap{$^c$} School of Physics, University of Edinburgh,
                    Edinburgh EH9 3JZ, UK \\
        \llap{$^d$} John von Neumann Institute NIC / DESY Zeuthen,
                    15738 Zeuthen, Germany \\
        \llap{$^e$} Institut f\"ur Theoretische Physik,
                    Universit\"at Leipzig,
                    04109 Leipzig, Germany \\
        \llap{$^f$} Theoretical Physics Division,
                    Department of Mathematical Sciences,
                    University of Liverpool,
                    Liverpool L69 3BX, UK \\
        \llap{$^g$} Department of Physics and Astronomy,
                    University of Kentucky,
                    Lexington KY 40506, USA \\
        \llap{$^h$} Konrad-Zuse-Zentrum f\"ur Informationstechnik Berlin,
                    14195 Berlin, Germany \\
        E-mail: \email{rhorsley@ph.ed.ac.uk} }

\author{QCDSF--UKQCD Collaboration}

\abstract{Results are presented for the lowest moment of the distribution
          amplitude for the $K^*$ vector meson. Both longitudinal
          and transverse moments are investigated. We use two flavours
          of $O(a)$ improved Wilson fermions, together with a non-perturbative
          renormalisation of the matrix element.}

\FullConference{The XXV International Symposium on Lattice Field Theory\\
		 July 30-4 August 2007\\
		 Regensburg, Germany}

\begin{document}


\section{Introduction}


`Rare decays' of $B$ mesons, such as $B \to K^*\gamma$,
$B \to K^* \mu^+ \mu^-$, $B \to \rho\gamma$, $\ldots$ where $b \to s\gamma$
are flavour changing neutral current or FCNC processes and
are thus not allowed at tree level by the GIM mechanism. However this makes
them sensitive to higher scales, and may affect various CKM matrix elements,
such as $V_{ts}$ or $V_{td}$. These exclusive events can be 
investigated at the LHC by the LHCb experiment. A theoretical framework
is provided by QCD factorisation,  eg, \cite{beneke99a,beneke01a},
(which is a heavy quark expansion in $1/m_b$), perturbative QCD
\cite{keum00a}, soft-collinear effective theory \cite{bauer00a}
or light-cone sum rules \cite{ball98a}. These give a decay amplitude
related to vector distribution amplitudes or vector DAs. These are usually
defined in the $\overline{MS}$ scheme at some scale $\mu$. In this article
we compute using lattice QCD the lowest moment of the $K^*$ DA.
Analogous computations have recently been performed for the spin $0$
particles $\pi$ and $K$, \cite{braun06a,boyle06a}.

As we have vector particles, with a polarisation vector, we have
two distinct DAs: $\phi^{\parallel}(\xi)$ and $\phi^\perp(\xi)$.
These are functions of $\xi \in [-1,+1]$, where $x = \half ( 1 + \xi)$ and
$1-x = \half ( 1 - \xi)$ are the fractions of meson momentum carried
by the quark and anti-quark respectively (in the infinite momentum frame).
An expansion in terms of Gegenbauer polynomials
\begin{eqnarray}
   \phi^{\parallel,\perp}(\xi)
      = \phi^{asymp}(\xi)
           \left(
              1 + \sum_1^\infty a_n^{\parallel,\perp}(\mu) C_n^{3/2}(\xi)
           \right) \,,
                                                       \nonumber
\end{eqnarray}
with
\begin{eqnarray}
   \phi^{asymp}(\xi) = { 3 \over 4} \left( 1 - \xi^2 \right) \,,
                                                       \nonumber
\end{eqnarray}
allows (possible) reconstruction of the full DA. In particular as
$a_n^{\parallel,\perp} \to 0$ when $\mu \to \infty$, we might hope
that knowledge of the lowest lowest few $a_n^{\parallel,\perp}$
coefficients suffices. Indeed the lattice computation is only capable
of giving low moments of DAs, defined by
\begin{eqnarray}
   \langle \xi^n \rangle^{\parallel, \perp}
      = \int_{-1}^{1} d\xi \xi^n \phi^{\parallel, \perp} (\xi, \mu) \,,
                                                       \nonumber
\end{eqnarray}
where $a^{\parallel,\perp}_1 = 5/3 \langle \xi \rangle^{\parallel,\perp}$,
$a^{\parallel,\perp}_2 = 7/12 (5\langle \xi^2 \rangle^{\parallel,\perp} - 1)$,
$\ldots$\, . As Gegenbauer polynomials are orthogonal polynomials with weight
$1-\xi^2$ and as $C_0^{3/2} = 1$ then the normalisation is such that
$\langle 1 \rangle^{\parallel, \perp} = 1$. Finally we note that
$G$-parity restricts the functional form of $\phi^{\parallel,\perp}_\rho$
to an even function of $\xi$ and so non-zero moments are
$\langle \xi \rangle_{K^*}$, $\langle \xi^2 \rangle_{K^*}$,
$\langle \xi^2 \rangle_{\rho}$, $\ldots$.


\section{Minkowski matrix elements}


Longitudinal matrix elements are given by
\begin{eqnarray}
   {\cal S}_{\mu_0\mu_1 \cdots \mu_n}
       \langle 0 | \widehat{\cal O}^{\M\mu_0\mu_1 \cdots \mu_n}
                 | V, \vec{p},\lambda \rangle
            = i m_V F^\parallel_V {\cal S}_{\mu_0\mu_1 \cdots \mu_n}
                \left[ \epsilon_\lambda^{\M\mu_0}
                         p^{\M\mu_1} p^{\M\mu_1} \cdots p^{\M\mu_n}
                \right] \langle \xi^n \rangle^{\parallel} \,,
                                                               \nonumber
\end{eqnarray}
with
\begin{eqnarray}
   {\cal O}^{\M\mu_0\mu_1 \cdots \mu_n}
               = i^n \overline{q} \gamma^{\M\mu_0}
                    \Dlr^{\M\mu_1}\Dlr^{\M\mu_2}\cdots\Dlr^{\M\mu_n} u \,,
                                                               \nonumber
\end{eqnarray}
where $q  = d$ or $s$, ${\cal S}$ means symmetrised and traceless
in these indices, $\Dlr = \Dr - \Dl$ and $\lambda$ is the polarisation
index. Correspondingly transverse matrix elements are given by
\begin{eqnarray}
   \lefteqn{{\cal S}_{\mu_0\mu_1 \cdots \mu_n}
         \langle 0 | \widehat{\cal O}^{\M\nu\mu_0\mu_1 \cdots \mu_n}
                   | V, \vec{p},\lambda \rangle =}
     & &                                                       \nonumber \\
     & & \hspace*{1.0in} i F_V^{\perp}{\cal S}_{\mu_0\mu_1 \cdots \mu_n}
                \left[ ( \epsilon_\lambda^{\M\nu} p^{\M\mu_0} -
                       \epsilon_\lambda^{\M\mu_0} p^{\M\nu} )
                       p^{\M\mu_1} \cdots  p^{\M\mu_n} 
                \right] \langle \xi^n \rangle^{\perp} \,,
                                                               \nonumber
\end{eqnarray}
(where
$\sigma^{\M\mu\nu} = \half \left[\gamma^{\M\mu},\gamma^{\M\nu} \right]$)
with operators
\begin{eqnarray}
   {\cal O}^{\M\nu\mu_0\mu_1 \cdots \mu_n}
     = i^n \overline{q} \sigma^{\M\nu\mu_0}
            \Dlr^{\M\mu_1}\Dlr^{\M\mu_2}\cdots\Dlr^{\M\mu_n} u \,.
                                                               \nonumber
\end{eqnarray}
This all looks rather complicated, but for no derivatives ($n=0$) the
equations reduce to the familar ones for the $F_V^{\parallel,\perp}$
decay constants, namely
\begin{eqnarray}
   \langle 0 | \widehat{V}^{\M\mu_0}
             | V, \vec{p},\lambda \rangle 
     = i m_V F^\parallel_V \epsilon_\lambda^{\M\mu_0} \,,
   \qquad V^{\M\mu_0} = \overline{q} \gamma^{\M\mu_0} u \,,
                                                               \nonumber
\end{eqnarray}
and
\begin{eqnarray}
   \langle 0 | \widehat{T}^{\M\nu\mu_0}
             | V, \vec{p},\lambda \rangle
     = i F_V^{\perp} ( \epsilon_\lambda^{\M\nu} p^{\M\mu_0} - 
                       \epsilon_\lambda^{\M\mu_0} p^{\M\nu} ) \,,
   \qquad T^{\M\nu\mu_0} = \overline{q} \sigma^{\M\nu\mu_0} u \,.
                                                               \nonumber
\end{eqnarray}
Thus we see that these equations have been normalised with
$F^{\parallel,\perp}_V$ to ensure, as required, that
$\langle 1 \rangle^{\parallel,\perp} =1$.


\section{The Lattice}


On the lattice we need a careful choice of lattice operators to avoid
mixing with same dimension operators, and worse mixing with lower dimensional
operators when $1/a$ subtractions are required. We shall consider only
$n=1$ operators here, the list \cite{gockeler06a} used is
\begin{eqnarray}
   \begin{tabular}{l|l|l}
      $n$  & Operator                     & Representation                \\
      \hline 
      $1$  &  ${\cal O}_i^{\parallel,a}
                          = {\cal O}_{\{i4\}}$ & $\tau_3^{(6)},   \, C=+$ \\
      $1$  &  ${\cal O}^{\parallel,b}
                          = {\cal O}_{44} - 
               {1 \over 3}\sum_i {\cal O}_{ii}$  & $\tau_1^{(3)}, \, C=+$ \\
      \hline 
   \end{tabular}
                                                               \nonumber
\end{eqnarray}
for the longitudinal operators, where
${\cal O}_{\mu_0\mu_1 \cdots \mu_n} = \overline{q} \gamma_{\mu_0}
\Dlr_{\mu_1}\Dlr_{\mu_2}\cdots\Dlr_{\mu_n} u$ and
\begin{eqnarray}
   \begin{tabular}{l|l|l}
      $n$  & Operator                     & Rep.                       \\
      \hline 
      $1$  & ${\cal O}^{\perp,a}_{ij} = {\cal O}_{ij4} + {\cal O}_{i4j} -
              {\cal O}_{4ij} - {\cal O}_{4ji}$, $i \not= j$
                                          & $\tau_2^{(8)}, \, C=+$      \\
      $1$  & ${\cal O}^{\perp,b}_i = {\cal O}_{i44} - 
              {1 \over 2} \sum_j {\cal O}_{ijj}$
                                          & $\tau_1^{(8)}, \, C=+$      \\
      \hline
   \end{tabular}
                                                               \nonumber
\end{eqnarray}
for the transverse operators, where ${\cal O}_{\nu\mu_0\mu_1 \cdots \mu_n}
= \overline{q} \gamma_\nu \gamma_{\mu_0}
\Dlr_{\mu_1}\Dlr_{\mu_2}\cdots\Dlr_{\mu_n} u$ ($\nu \not= {\mu_0}$).
The operators belonging to different (hypercubical) representations
have been labelled by `a' and `b', and should give the same results,
at least in the continuum limit. (Further results,
including $n=2$ operators will appear in \cite{gockeler07a}.)

Correlation functions are then defined, where
\begin{eqnarray}
   C_{{\cal O}\Omega}(t;\vec{p})
      = \langle \widehat{\cal O}(t;\vec{p})
                \widehat{\Omega}(0;\vec{p})^{\dagger} \rangle \,,
                                                               \nonumber
\end{eqnarray}
with $\Omega = V$ or $T$, where to improve the signal these operators
have been `Jacobi' smeared. Then inserting complete sets of states
in the standard way gives correlation functions involving
$\langle 0| \widehat{\Omega} |V, \vec{p},\lambda\rangle$ and
$\langle 0| \widehat{\cal O} |V, \vec{p},\lambda\rangle$. The unwanted
$\langle 0| \widehat{\Omega} |V, \vec{p},\lambda\rangle$ may be cancelled
by forming ratios. For example we find for some of the (bare) operators
\begin{itemize}

   \item Longitudinal
         \begin{eqnarray}
            { \third \sum_i C_{{\cal O}^{\parallel,a}_iV_i}(t;\vec{p}) \over
              \third \sum_i C_{V_iV_i}(t;\vec{p}) }
                 &=& - {1 \over 2} E_V 
                     \left( { 2E_V^2 + m_V^2 \over
                     E_V^2 + 2m_V^2 } \right)
                      \, \tanh E_V(\half N_T - t) \, 
                       \langle \xi \rangle^{\parallel}_a
                                                  \nonumber \\
            { C_{{\cal O}^{\parallel,b}V_i}(t;\vec{p}) \over
              \third \sum_i C_{V_iV_i}(t;\vec{p}) }
                 &=& - {4 \over 3} i p_i
                     \left( { 3 E_V^2 \over E_V^2 + 2m_V^2 } \right)
                      \, \langle \xi \rangle^{\parallel}_b
                                                  \nonumber
         \end{eqnarray}

   \item Transverse
         \begin{eqnarray}
            { C_{{\cal O}^{\perp,a}_{lm}V_n}(t;\vec{p}) \over
              \third \sum_i C_{T_iV_i}(t;\vec{p}) }
                &=& 3 i \delta_{ln} p_m
                  \, \langle \xi \rangle^\perp_a 
                                                  \nonumber \\
            { \third \sum_i C_{{\cal O}^{\perp,b}_iV_i}(t;\vec{p}) \over
              \third \sum_i C_{T_iV_i}(t;\vec{p}) }
                &=& - E_V 
                    \left( { 4E_V^2 - m_V^2 \over 3 E_V^2 } \right)
                      \, \coth E_V(\half N_T - t) \, 
                       \langle \xi \rangle^\perp_b
                                                  \nonumber
         \end{eqnarray}

\end{itemize}
and similar expressions for the other operators as
$\Omega$ (in the above $V$) can also be replaced by $T$
giving further ratios. Thus many cross checks are possible.
Note that the $t$ fit function is known and may be either
$\tanh$, $\coth$ or $1$. Also half the $n=1$ operators can be
measured at zero momentum; the others cannot. However for those
operators a non-zero ratio requires only a single unit of momentum
in one direction. We choose the lowest possible momentum,
$|\vec{p}| = 2\pi/N_S$ and average over the three spatial directions.

We use unquenched $n_f=2$, $O(a)$ improved clover fermions in our
simulations, the lattices employed being:
\begin{table}[h]
   \begin{center}
      \begin{tabular}{||l|l|c|l||c|c|c|c|c||}
         \hline
          $\beta$ & $\kappa_{sea}$ & $N_S^3 \times N_T$ & Trajs &
          $m_{ps}/m_V$ & $m_{ps}L_S$ & $a[\mbox{fm}]$ & $L_S[\mbox{fm}]$  &
           $m_{ps}[\mbox{MeV}]$ \\
         \hline
         \hline

 5.29 & 0.1350 & $16^3\times 32$ & 5700 &
 0.76 & 6.7 & 0.075 & 1.20 & 1100 \\

 5.29 & 0.1355 & $24^3\times 48$ & 2100 &
 0.70 & 7.8 & 0.075 & 1.81 & 860 \\

 5.29 & 0.1359 & $24^3\times 48$ & 4900 &
 0.62 & 5.7 & 0.075 & 1.81 & 630 \\
         \hline
      \end{tabular}
   \end{center}
\end{table}
\newline
together with various $\kappa_{val}$ for the valence quarks.
Note that $L_S = a N_S$ and $m_{\pi^+}/m_{\rho^+} \sim 0.18$.
The scale is set from the $r_0$ force scale, using a value of
of $r_0 = 0.467\,\mbox{fm} \equiv 1/422.5\,\mbox{MeV}$. $a$ is determined
from extrapolating $(r_0/a)$ to the chiral limit
(presently giving $(r_0/a)_c(\beta = 5.29) = 6.20(3)$).
No operator improvement has been attempted, although experience
from quenched unpolarised operators has indicated that these
effects are probably small, \cite{gockeler04a}.

A non-perturbative renormalisation -- $RI^{\prime}-MOM$ method has been
used to determine the renormalisation constants.
($Z^{\rimom}$ is computed numerically and from this $Z^{\rgi}$ is determined.
This is then converted to $Z^{\msbar}(\mu = 2\,\mbox{GeV})$, which is the
scheme and scale that all our results are presented here.) For more
details see the forthcoming paper \cite{gockeler07b}.

\begin{figure}[t]
   \hspace{1.25in}
   \epsfxsize=7.00cm
      \epsfbox{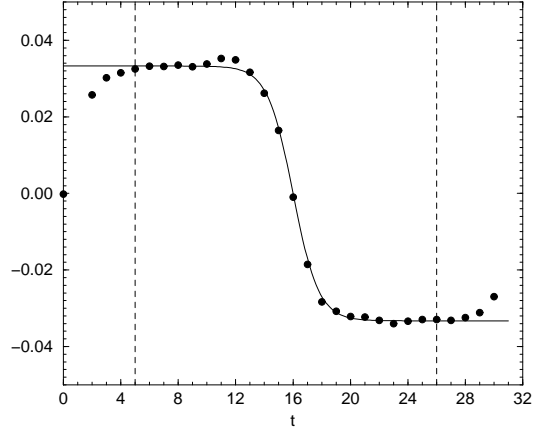}
   \caption{The ratio
            $- {2 \over m_V} \third \sum_i C_{{\cal O}^a_iV_i}(t;\vec{0}) /
             \third \sum_i C_{V_iV_i}(t;\vec{0}) \equiv
            \langle \xi \rangle^{\parallel}_a \tanh m_V(\half N_T -t)$
            versus $t$ for $\beta = 5.29$, $\kappa_{sea}=0.1350$,
            $\kappa_{val}=(0.1355,0.1343)$, $\vec{p}=\vec{0}$.
            The results are denoted by filled circles. Also shown is a
            one parameter fit (the $m_V$ mass having been
            determined previously). The fit range is denoted by vertical
            dashed lines. The operator has been renormalised to
            the $\overline{MS}$ scheme at $\mu = 2\, \mbox{GeV}$,
            so that the value obtained corresponds directly to a point
            in Fig.~\protect\ref{fig_xi1_mks2-mps2_V_1pic_070727_lat07_writeup}
            (the sixth point from the left).}
   \label{fig_rda1a_gi_kp13550kp13430_peq0+fit_lat07_writeup}
\end{figure}

A (typical) result for the ratio is shown in
Fig.~\ref{fig_rda1a_gi_kp13550kp13430_peq0+fit_lat07_writeup},
where we observe a clear $\tanh$ function.


\section{Results}


As noted previously, {\it odd} moments vanish for degenerate mass (valence)
quarks and thus we have ($m_{q_2} < m_{q_1}$)
\begin{eqnarray}
   \langle \xi \rangle^{\parallel,\perp}
         &\propto& m_{q_1} - m_{q_2}  \propto m_{q_1} + m_{q_2} - 2m_{q_2}
                                                  \nonumber \\
         &\propto& m_{Kps}^2 - m_{ps}^2 \,,
                                                  \nonumber
\end{eqnarray}
where $m_{ps}$ is a pseudoscalar meson with degenerate mass quarks
and $m_{Kps}$ is a pseudoscalar meson with possibly non-degenerate
mass quarks. (For the {\it even} moments, not considered here,
there is no such restriction and are just symmetric
in the quark masses.) For $\langle \xi \rangle^{\parallel,\perp}_{K^*}$
we first, for fixed $m_{sea}$, plot $\langle \xi \rangle^{\parallel,\perp}$
against (valence pseudoscalar masses) $m_{Kps}^2 - m_{ps}^2$
and interpolate to the physical point $m_K^2 - m_{\pi}^2$, \cite{braun06a}.
This is then taken as a function of $m_{sea} \propto m_{ps}^2$
and extrapolated to the chiral limit to give finally
$\langle \xi \rangle^{\parallel,\perp}_{K^*}$.

\begin{figure}[p]
   \hspace{1.00in}
   \epsfxsize=9.00cm
      \epsfbox{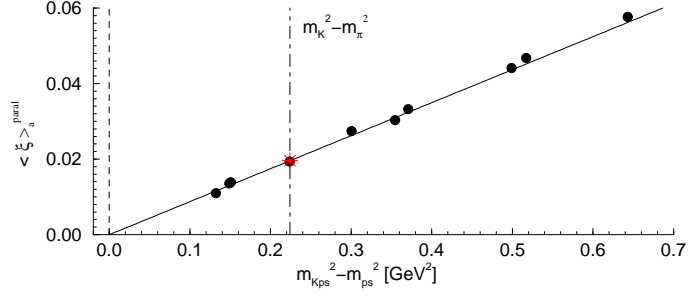}
   \caption{$\langle \xi \rangle_a^{\parallel}$ versus $m_{Kps}^2 - m_{ps}^2$
            in the $\overline{MS}$ scheme at $\mu = 2\,\mbox{GeV}$
            for $\beta = 5.9$, $\kappa_{sea}=0.1350$, $\vec{p}=\vec{0}$
            for various valence quark combinations. A linear fit vanishing
            when the two valence quark masses are the same is also shown.
            The red star shows the value when
            $m_{Kps}^2-m_{ps}^2 = m_K^2 - m_\pi^2$.}
   \label{fig_xi1_mks2-mps2_V_1pic_070727_lat07_writeup}
\end{figure}

\begin{figure}[p]
   \hspace{1.00in}
   \epsfxsize=9.00cm
      \epsfbox{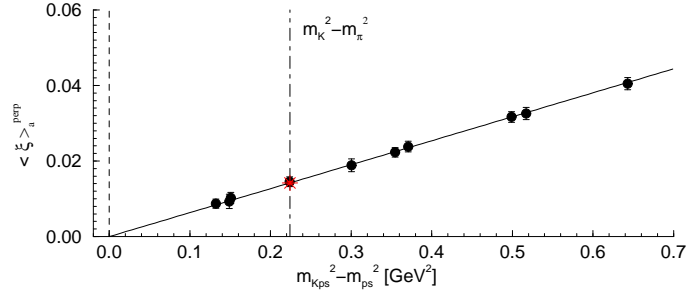}
   \caption{$\langle \xi \rangle_a^{\perp}$ versus $m_{Kps}^2 - m_{ps}^2$.
            Same notation as for
            Fig.~\protect\ref{fig_xi1_mks2-mps2_V_1pic_070727_lat07_writeup}.
            Note that here we work at finite momentum $|\vec{p}| = 2\pi/N_S$.}
   \label{fig_xi1perp_mks2-mps2_V_1pic_070727_lat07_writeup}
\end{figure}

In Fig.~\ref{fig_xi1_mks2-mps2_V_1pic_070727_lat07_writeup}
we show $\langle \xi \rangle_a^{\parallel}$ versus $m_{Kps}^2 - m_{ps}^2$
together with a one-parameter fit passing through the origin. Also shown
(red star) is the value when $m_{Kps}^2-m_{ps}^2 = m_K^2 - m_\pi^2$.
Fig.~\ref{fig_xi1perp_mks2-mps2_V_1pic_070727_lat07_writeup}
shows the corresponding results for $\langle \xi \rangle_a^{\perp}$.

\begin{figure}[p]
   \hspace{1.00in}
   \epsfxsize=9.00cm
      \epsfbox{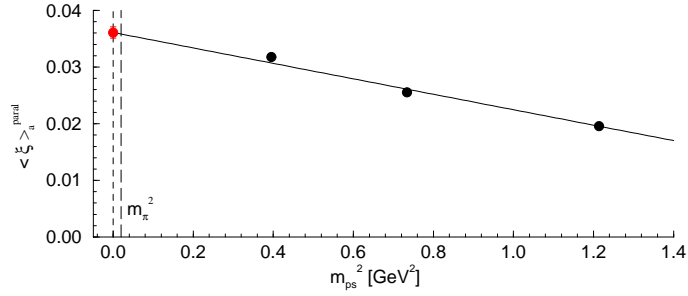}
   \caption{$\langle \xi \rangle_a^{\parallel}$ versus $m_{ps}^2$
            for the three sea quark masses $\kappa_{sea} = 0.1350$, 
            $0.1355$ and $0.1359$ (black circles), together with a linear
            extrapolation to the chiral limit (red circle).} 
   \label{fig_xi1_mps2r0cGeV2_b5p29_1pic_070728_lat07_writeup}
\end{figure}

\begin{figure}[p]
   \hspace{1.00in}
   \epsfxsize=9.00cm
      \epsfbox{
        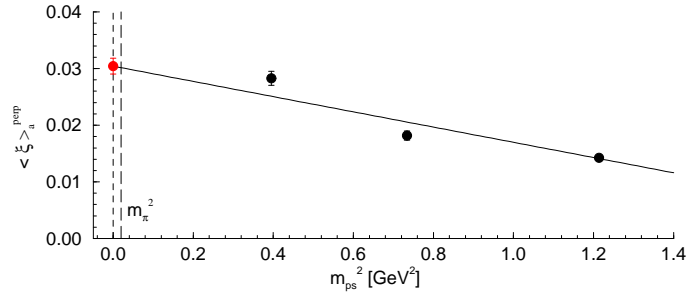}
   \caption{$\langle \xi \rangle_a^{\perp}$ versus $m_{ps}^2$.
            Same notation as in Fig.~\protect
            \ref{fig_xi1_mps2r0cGeV2_b5p29_1pic_070728_lat07_writeup}.}
   \label{fig_xi1trans_mps2r0cGeV2_b5p29_1pic_070728_lat07_writeup}
\end{figure}

As discussed previously we must now extrapolate $m_{sea} \propto m_{ps}^2$
to the chiral limit (the difference between this and $m_\pi^2$ is
negligible). In Fig.~\ref{fig_xi1_mps2r0cGeV2_b5p29_1pic_070728_lat07_writeup}
we show this extrapolation for $\langle \xi \rangle_a^{\parallel}$ giving
an estimate for $\langle \xi \rangle_{K^*}^{\parallel}$.
In Fig.~\ref{fig_xi1trans_mps2r0cGeV2_b5p29_1pic_070728_lat07_writeup}
we show the equivalent picture for $\langle \xi \rangle_a^{\perp}$ leading
to a value for $\langle \xi \rangle_{K^*}^{\perp}$.

This is repeated for other channels and we thus finally arrive at the
(preliminary) results
\begin{eqnarray}
   \begin{array}{lll}
      \langle \xi \rangle^{\parallel}_{K^*}    &\approx& 0.033(2)(4) \\
      \langle \xi \rangle^{\perp}_{K^*}        &\approx& 0.030(2)(8) \\
   \end{array}
   \quad \mbox{or} \quad
   \begin{array}{lll}
      a_{1K^*}^{\parallel}    &\approx& 0.055(3)(7) \\
      a_{1K^*}^{\perp}        &\approx& 0.050(3)(13)\\
   \end{array}
                                                  \nonumber
\end{eqnarray}
(in the $\overline{MS}$-scheme at a scale of $\mu = 2\,\mbox{GeV}$)
where the first error comes from the spread of channels presently analysed
and the second error is an estimation of possible chiral extrapolation 
error (the fit being repeated dropping one data point). Also
any discretisation errors have been ignored.

These are to be compared with the results from sum rule estimates
of $a_{1K^*}^{\parallel} \approx 0.02(2)$,
$a_{1K^*}^{\perp} \approx 0.03(3)$ \cite{ball07a} at the same scale,
 and the limit function $\phi^{asymp}(\xi)$ giving
$a_{1K^*}^{\parallel,\perp} = 0$.
Potentially lattice results are more reliable than sum rule
estimates and may help in a reconstruction of the vector distribution
amplitude.

Our conclusion is that a lattice determination of (moments of) vector DAs
is possible. We plan to extend these results to
lighter pseudoscalar masses, $\beta = 5.40$ (a finer lattice) and to
$\langle \xi^2 \rangle^{\parallel,\perp}$ for both the $K^*$ and $\rho$.
Further results (including the zero moment decay constants) will appear in
\cite{gockeler07a}.


\section*{Acknowledgements}


The numerical calculations have been performed on the Hitachi SR8000 at
LRZ (Munich), on the Cray T3E at NIC (J\"ulich) and ZIB (Berlin),
as well as on the APEmille and APEnext at DESY (Zeuthen),
and on the BlueGeneLs at NIC/J\"ulich, EPCC
at Edinburgh and KEK at Tsukuba by the Kanazawa group as part of the
DIK research programme. We thank all institutions.
This work has been supported in part by
the EU Integrated Infrastructure Initiative Hadron Physics (I3HP) under
contract RII3-CT-2004-506078 and by the DFG under contract
FOR 465 (Forschergruppe Gitter-Hadronen-Ph\"anomenologie).



\end{document}